\documentclass[12pt]{iopart}
\pdfoutput=1
\usepackage{graphicx}
\usepackage[latin1]{inputenc}
\usepackage{amsfonts}
\usepackage{amssymb}
\graphicspath{{./images/}}
\begin{document}
\title[Rounding metabolic networks]{Uniform sampling of steady states in metabolic networks: heterogeneous scales and rounding.}
\author{D.De Martino$^{1}$, M.Mori$^{2}$, V.Parisi$^{2}$}
\address{$^1$ Center for life nanoscience CLNS-IIT, P.le A.Moro 2, 00815, Rome, Italy}
\address{$^2$ Dipartimento di Fisica, Sapienza Universit\'{a} di Roma, P.le A.Moro 2, 00815, Rome, Italy}

\begin{abstract}
The uniform sampling of convex polytopes is an interesting computational problem with many applications in inference from linear constraints, but the performances of sampling algorithms can be affected by ill-conditioning. This is the case of inferring the feasible steady states in models of metabolic networks, since they can show heterogeneous time scales . In this work we focus on rounding procedures based  on building an ellipsoid that closely matches the sampling space, that can be used to define an efficient { hit--and--run (HR)} Markov Chain Monte Carlo. In this way the uniformity of the sampling { of the convex space of interest is rigorously guaranteed,} at odds with non markovian methods. We analyze and compare three rounding methods in order to sample the feasible steady states of metabolic networks of three models of growing size { up to} genomic scale. The first is based on principal component analysis (PCA), the second on linear programming (LP) and finally we employ the lovasz ellipsoid method (LEM).
Our results show that a rounding procedure is mandatory for the application of the { HR} in these inference problem and suggest that a combination of LEM or LP with a subsequent PCA perform the best.   We finally compare the distributions of the HR with that of two heuristics based on the  Artificially Centered hit--and--run (ACHR), {\em gpSampler } and {\em optGpSampler}. They show a good agreement with the results of the HR for the small network, while on genome scale models present inconsistencies.
\end{abstract}

\maketitle

\section{Introduction}
The metabolism of cells is based on a complex metabolic network of chemical reactions performed by enzymes{, which} are 
able to degrade nutrients {in order to produce biomass and generate the energy needed to sustain all other tasks the cell has to perform\cite{biochemlen}.} The high-throughput data coming from genome sequencing of single organisms can be used to reconstruct the complete set of enzymes devoted to metabolic functions, leading to models of metabolism at the scale of the whole genome,  
whose analysis is computationally challenging\cite{palssonbook}. 
If we want to model a metabolic  system in terms of the dynamics of the concentration levels, even upon assuming well-mixing (no space) and neglecting noise (continuum limit), we have a very large non-linear dynamical system whose parameters are mostly unknown.  
For a chemical reaction  network  in which $M$ metabolites participate in $N$ reactions (where $N,M \simeq \mathcal O(10^{2-3})$ in genome-scale models) with the  stoichiometry encoded in a matrix $\mathbf{S}=\{S_{\mu r}\}$, 
the concentrations $c_\mu$ change in time according to mass-balance equations
\begin{equation}\label{eq1}
\dot{\mathbf{c}} = \mathbf{S \cdot v}
\end{equation}
where $v_i$ is the flux of the reaction $i$ that in turn is a possibly unknown  function of the concentration levels  $v_i(\mathbf{c})$, with possibly unknown parameters. 
A simplifying hypothesis is to assume the system in a steady state $\dot{\mathbf{c}}=0$. 
The fluxes are further bounded in certain ranges $v_r \in [v_{r}^{{\rm min}},v_{r}^{{\rm max}}]$ that take into account thermodynamical irreversibility, kinetic limits and physiological constraints.
The set of constraints
\begin{eqnarray}\label{eq3}
\mathbf{S \cdot v}=0, \nonumber \\
v_r \in [v_{r}^{{\rm min}},v_{r}^{{\rm max}}]
\end{eqnarray}  
defines  a convex closed set in the space of reaction fluxes: a polytope from which feasible steady states should be inferred.     
 
In general, the problem of the uniform sampling of convex bodies in high dimensions is both of theoretical and practical importance{.}
From a theoretical viewpoint it leads to  polynomial-time approximate algorithms for the calculation of the volume of a convex body\cite{Simonovits03}, 
whose exact determination is a \#P-hard problem\cite{Dyer-Frieze}.
On the other hand general problems of inference from linear constraints require an uniform sampling of the points inside a convex polytope: 
other examples apart from metabolic network analysis\cite{schelljbc} include compressed sensing\cite{comprsens}, freezing transition of hard spheres\cite{freezingHR} and density reconstruction from gravitational lensing in astrophysics\cite{lubini}. 
The knowledge of all the vertices characterizes completely a polytope but deterministic algorithms that perform an exhaustive enumeration can be infeasible  in high dimensions since the number of such vertices could scale exponentially with the dimension. An alternative is to carry out a statistical analysis of the space by means of Monte Carlo methods\cite{montecarlokrauth}.
A static approach with a simple rejection rule is feasible for low dimensions\cite{MC2} but we have to recur to dynamical methods in high dimensions. The faster and most popular algorithm in order to sample points inside convex bodies is the { hit--and--run} (HR) Markov Chain Monte Carlo\cite{HR,HRb}. 

The  mixing time of the { HR} scales as a polynomial of  the dimensions  of the body  but the method can suffer of ill{--}conditioning if the body is highly heterogeneous. More precisely the mixing time $\tau$ scales like\cite{lovasz1}
\begin{equation}
\tau \simeq \mathcal O(D^2 R^2/r^2)
\end{equation} 
where $D$ is the dimension of the polytope, $R,r$ are the radii  of respectively the  minimum inscribing  and the maximum inscribed balls: $R/r$ has been called the sandwitching ratio of the body. The sandwitching ratio quantifies the degree of ill{--}conditioning of the sampling problem  and we will refer to it as its condition number.
  
Several alternatives have been proposed to the { HR} dynamics. A simple one consists in a rather coarse approximation: a certain number of vertices is calculated by linear programming applied to random linear objective functions and the points inside can be sampled by interpolation. This approximation suffers from the fact that we are neglecting possibly an exponentially large number of vertices{,} and it has been shown that this  leads to wrong results even for simple hypercubes\cite{lubini}. { Artificially Centered hit--and--run (ACHR)\cite{kaufsmith}, is} a non-markovian modification of the { HR} algorithm by selecting the sample directions along previously sampled points. 
{ ACHR has} been widely used in order to sample flux configurations in metabolic networks \cite{gpsampler, 2014optGpSampler, bordbar} but it has the drawback that its non-markovian nature doesn't guarantee the convergence to an uniform distribution.
{ Finally, } the sampling problem has been reformulated within the { framework of Message Passing (MP)  algorithms\cite{Braunstein, MPperez}, which allow very fast sampling, but work under the approximation of a tree-like network and 
are not guaranteed in general to converge to an uniform distribution.}
On the other hand it is known that the sandwitching ratio of a polytope can be reduced to at most 
$\sqrt D$ for centrally simmetric polytopes and to  $D$ in general, by an affine transformation defined by 
the so-called {Loewner--John} Ellipsoid\cite{Ball}, i.e. the ellipsoid of maximal volume contained in the polytope.
Unfortunatively this ellipsoid cannot be found in polynomial time, but it has been shown by {L.~lovasz} that a weaker form of the { Loewner--John} ellipsoid, with a factor of $D^{3/2}$, can be found in polynomial time\cite{lovaszbook} 
The feasible steady states of a metabolic network can show very heterogeneous scales on genome-scale models:
previous samplings\cite{almaas} seem to indicate that the distribution of flux scales can span 
$5$ orders of magnitudes, and we should thus expect $R/r \simeq 10^5$ in practical cases, that means that
the ill{--}conditioning is a crucial issue in this inference problem.
The focus of this work is on the reduction of the condition number in the uniform sampling of convex polytopes by finding an ellipsoid that closely matches the underlying space. 
We use this matching ellipsoid  to extract the direction of the { HR}, a procedure that is equivalent to an affine transformation and eliminates the ill{--}conditioning.
We will analyze and compare three methods: the first is based on building an ellipsoid by applying  principal component analysis (PCA) to previous samplings, the second, inspired by a technique called {{\em Flux Variability Analysis} (FVA)}, 
uses linear programming (LP) in order to calculate the axis of the ellipsoid by maximizing and minimizing repeatedly the constraints defining the polytope, and finally the lovasz ellipsoid method\cite{lovaszbook} (see {Fig.~1} for a sketch).
\begin{figure}[h!]
\begin{center}
\includegraphics*[width=.6\textwidth,angle=270]{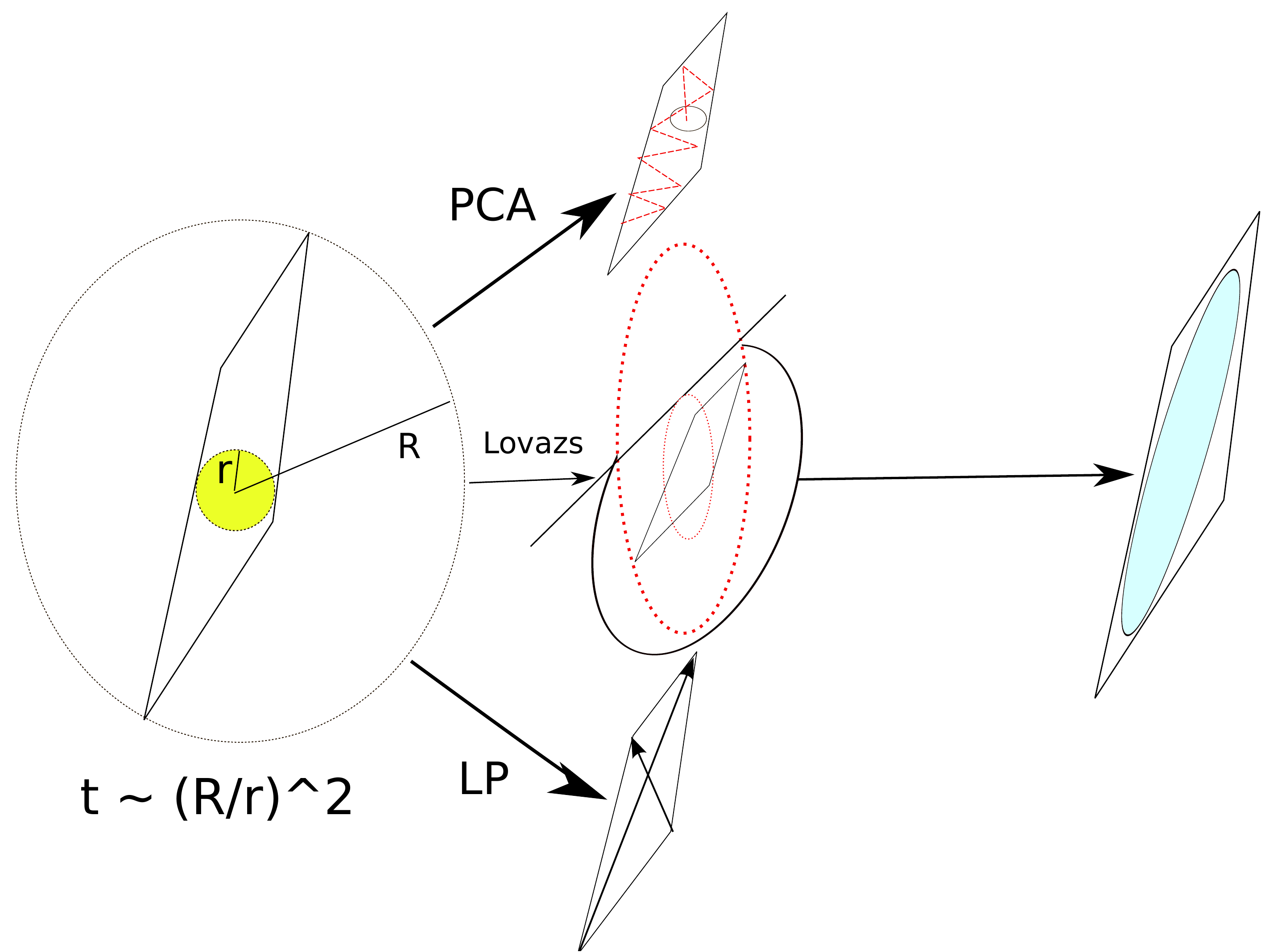}
\caption{Sketch of the ill{--}conditioning problem in the uniform sampling from the polytope of metabolic steady states. We propose to build a matching ellipsoid in three ways: PCA, LP and LEM: see section  Materials and methods for a full description.}
\end{center}
\end{figure}
We will focus on the problem of characterizing the space of feasible steady states in three networks of growing size: the catabolic core of the reconstruction of the metabolism of the bacterium { Escherichia~coli} iAF1260\cite{Feist} and two genome scale models, respectively of {\em Saccaromyces Cervisiae} (SCiND750) \cite{duarte2004reconstruction} and cervix squamous epithelial cells\cite{Recon2} (CERVIX).
We then compare our uniform sampling with the results provided by two ACHR-based heuristics provided with the COBRA toolbox {\em gpSampler}\cite{gpsampler} and {\em optGpSampler}\cite{2014optGpSampler}.
The description of the methods, i.e.  the construction of ellipsoids that matches the space by means of PCA, LP and lovasz method follows thereafter.
Finally we draw out some conclusions and perspectives.

\section{Results}
We first discuss  the application of the rounding methods in order to sample the feasible steady states of the {\em E. coli}'s metabolic network reconstruction iAF1260 catabolic core.
{ This a network with $M=72$ metabolites and $N=95$ reactions, including all exchange reactions}. We consider the flux bounds provided with the model and employ bounds for the exchange fluxes that include possibility to intake the main elements that are needed in order to produce biomass: glucose, oxygen, ammonia, water and phosphate. 
Upon deleting null reactions we are left with a network of { $M=68$ metabolites and $N=86$ reactions}.
The resulting polytope has $D=23$ dimensions. In Fig.~2 we report the integrated autocorrelation times of the fluxes during the { HR} with different pre-processing schedules, ordered for increasing values. The measure of integrated autocorrelation times is a rather standard procedure in order to asses the reliability of average estimates in Markov chains, we refer to the supplementary materials for further details.
\begin{figure}[h!]
\begin{center}
\includegraphics*[width=.6\textwidth]{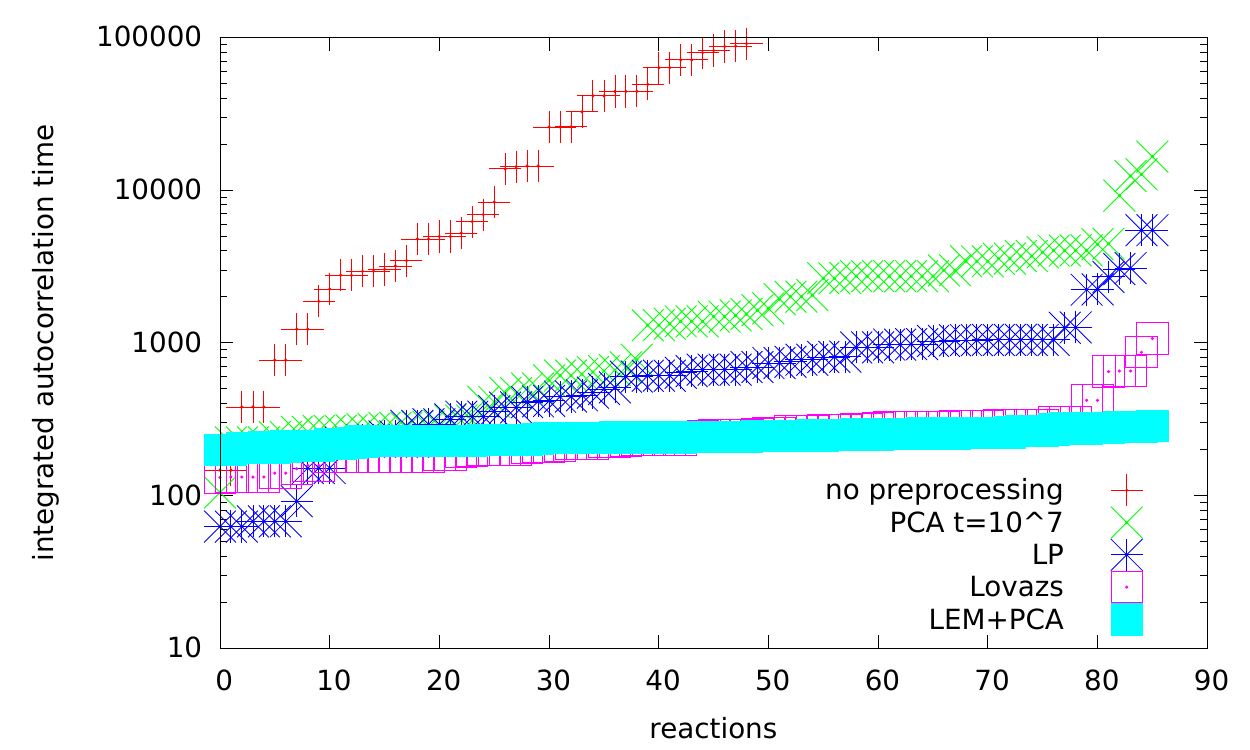}
\caption{Integrated autocorrelation times ordered for increasing values of the steady state fluxes of the core model { \em E.~coli} iAF1260 during an { hit--and--run} dynamics. Several preprocessing rounding procedures are compared: none, PCA, LP, LEM, LEM+PCA.}
\end{center}
\end{figure}
\begin{table*}[h!!!!!!]
\begin{tabular}{| c | c   | c | c |c  | c | }  
\hline 
Time  &  Normal  & PCA  &  LP & LEM & LEM(LP)+PCA  \\ 
\hline 
Preprocessing time (s)& $0$ & $40$ & { $7$} & $4$ & 4.2 ({ $7.2$}) \\
Max.int.autocor.time (mc steps) & $\mathcal O(10^9)$  & $1.6\cdot 10^4 $  & $5.4\cdot 10^3 $  & $1.1 \cdot 10^3 $ & $285 $ \\
\hline
\end{tabular}
\caption{Preprocessing time and maximum integrated autocorrelation time 
for the { hit--and--run} algorithms being examined on the {{\em E. coli}} core iAF1260 metabolic network. On an Intel dual core at $3.06 GHz$ using a single thread.}
\end{table*}
In { Table~1} we report the machine time needed to obtain the rounding ellipsoid and the measured maximum integrated autocorrelation time among the sampled reaction fluxes.
The times for the algorithm without preprocessing are very large, i.e. {$\mathcal{O}(10^9)$} Monte Carlo steps (hours for our implementation) in order to obtain reliable estimates for the flux averages. 
The preprocessing with PCA alone improves the situation,  but the attainment to stationarity of the covariance matrix is still lacking. The LP and LEM methods alone successfully reduce the condition number rendering the sampling possible in feasible computational times.
In particular the lovasz method performs better in this case.
Finally, the best result that minimizes the integrated autocorrelation times comes upon combining LEM method (or LP) with a subsequent PCA. In this way it is possible to obtain the ellipsoid directly from a good estimator of the stationary connected covariance matrix. 
{
Once the polytope has been rounded with a matching ellipsoid the mixing time of the  HR Markov chain scales as a polynomial of the system size and it would be possible to perform a rigorous uniform sampling even of genome scale network models, whose number of reactions is typically of $\mathcal{O}(10^3)$.
We have thus performed our preprocessing and subsequent sampling of two genome scale models, i.e. the model for  {\em Saccaromices Cerevisiae} SCiND750\cite{duarte2004reconstruction} and a model  of cervix squamous epithelial cells (CERVIX) from the reconstruction Recon 2\cite{Recon2}. 
We consider the first with default bounds for the uptakes while for the second we leave the network completely open to test the different cases. After removal of blocked reactions, the resulting dimensions of the polytope are $D=180$ and $D=694$, respectively. It turns out that with our implementation the more convenient rounding procedure consists in using the LP approach with a subsequent PCA. The procedure is intensive but feasible, it requires approximately $30$m to find an ellipsoid for SCiND750 and $3$h for CERVIX, for a sake of comparison the lovasz method requires $15$h for SCiND750. From the analysis of integrated autocorrelation times we get a maximum value in MC steps of $\mathcal{O}(10^4)$ where a MC step can performed in $\mathcal{O}(ms)$, as it is summarized in Table~2.
\begin{table*}[h!!!!!!]
\begin{centering}
\begin{tabular}{| c | c   | c | }  
\hline 
Time  & SciND750   & CERVIX    \\ 
\hline 
Preprocessing time (h)& 0.5 & 3 \\
Max.int.autocor.time (mc steps) & $1.6\cdot10^4$    & $7.4\cdot 10^4 $  \\
Average time for one mc step (ms) & 2 & 8 \\
\hline
\end{tabular}
\caption{Time performance of our implementation of the { hit--and--run} on genome scale networks. On an Intel dual core processor with clock rate $3.06 GHz$ using a single thread.}
\end{centering}
\end{table*}
We can characterize heterogeneity of scales by looking at the length of the diameters of the ellipsoid obtained by diagonalizing the covariance matrix{, which} we show in Fig.~3: even the small network spans across four order of magnitude, while the genome scale network SCiND750 spans across eight orders of magnitude. This strong heterogeneity would affect dramatically the performances of diffusive montecarlo markov chains without some pre-processing.
On the other hand, the largest CERVIX network, being completely open, it spans across three order of  magnitude in a continuous fashion.  
\begin{figure}[h!]
\begin{center}
\includegraphics*[width=.32\textwidth]{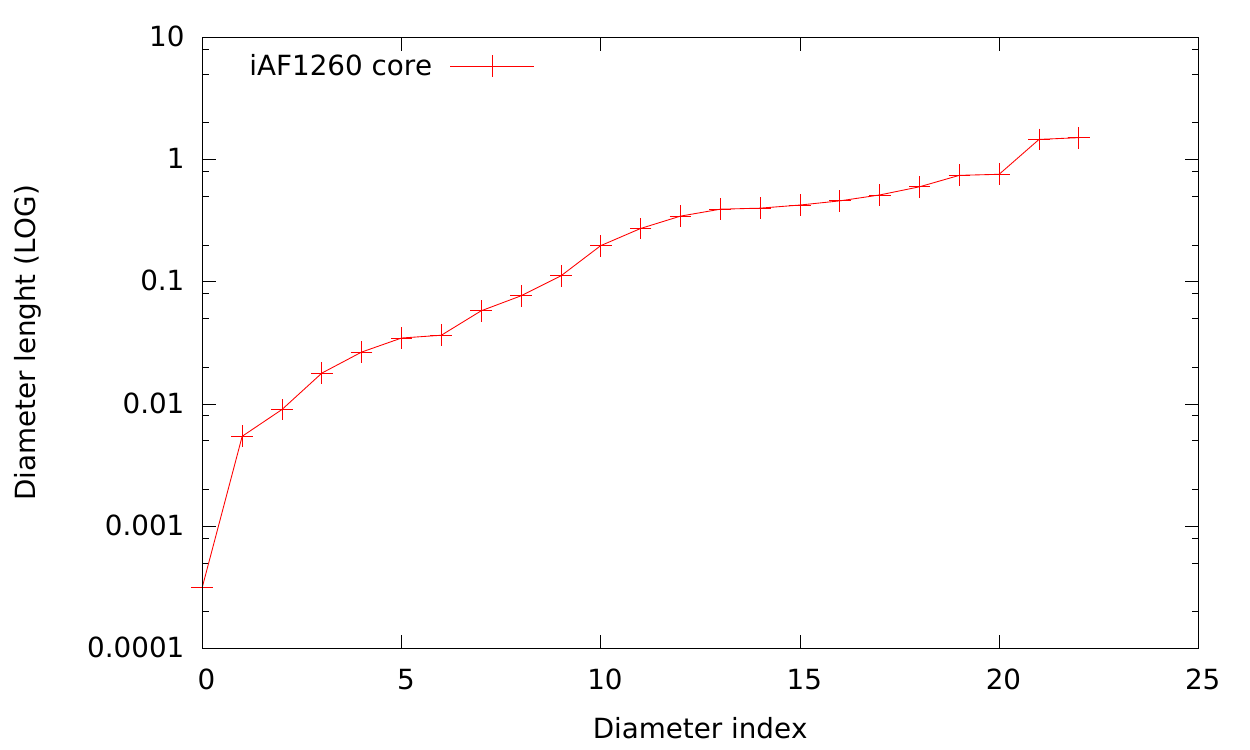}
\includegraphics*[width=.32\textwidth]{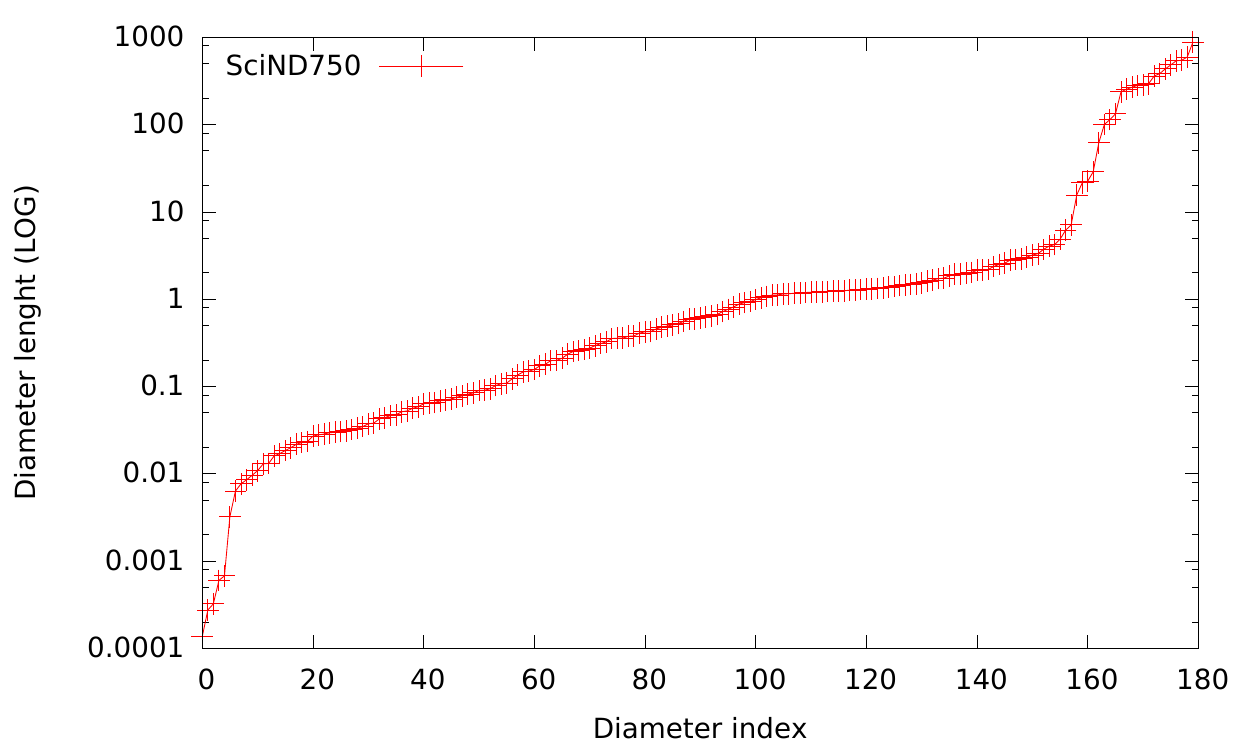}
\includegraphics*[width=.32\textwidth]{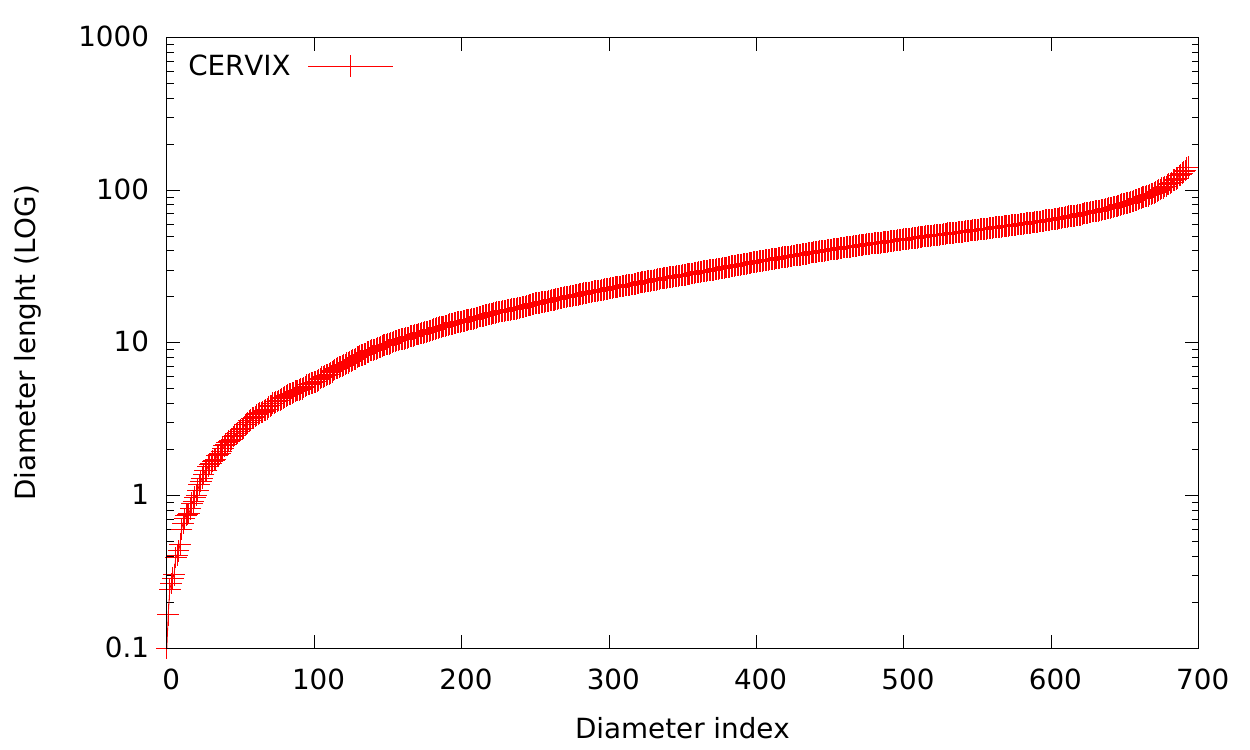}
\caption{Diameters, ordered for increasing length, of the ellipsoids retrieved from the models iAF1260 Core (left), SciND750 (center) and CERVIX (right). Ellipsoids are obtained from the diagonalization of the stationary connected covariance matrix calculated by combining LP and PCA.}
\end{center}
\end{figure}

}
{
\subsection{Comparison with artificially centered { hit--and--run} based {methods}}
We have thus seen that a rigorous uniform sampling of steady states in genome-scale metabolic networks is feasible, albeit intensive, with the { HR} algorithm once the ill{--}conditioning has been removed with a rounding procedure. We can use here our results to test the validity of ACHR-based heuristics, that can be used with the COBRA\cite{gpsampler} toolbox, {\em gpSampler}\cite{gpsampler} and {\em optgpSampler}\cite{2014optGpSampler}.  
We have checked that the two ACHR methods give very similar distributions upon waiting an effective convergence of {\em gpSampler}, that for the largest network analyzed requires around $1$ week of machine time on an Intel dual core at 3.06 GHz using a single thread.
We acknowledge on the other hand that {\em optGpSampler} converges
in much shorter times, and it is faster of the { HR} with our implementation  once the rounding preprocessing time is taken into account.
On the small { {\it E.~coli}} Core network half of the marginal flux distributions retrieved by ACHR based methods are consistent with the ones obtained by the { HR}
according to the Kolmogorov-Smirnov test (KS)\cite{kolmogorov1933sulla}  with a confidence of $5\%$.

The remaining marginal flux distributions are not in rigorous agreement but they provide a reliable approximation as it can be seen by the low values of the Kullback-Leibler divergence (KLD).
If we have two distributions $P(x)$ and $Q(x)$ the KLD is defined as
\begin{equation}
KLD(Q|P)=\int dx P(x) \log_2(P(x)/Q(x)),
\end{equation}
it is measured in {\em bits} and it quantifies the information that is lost by approximating $P$ with $Q$. More precisely, if we extract $N$ points from $Q$ we would be {\em deceived} with probability $2^{-N \cdot KLD(Q|P)}$, i.e. we would believe the points come from $P$\cite{mackay}. 
On the genome scale networks the marginal distributions retrieved by ACHR based methods do not pass the KS test, but they give an approximation. We have classified the level of 
approximation according to the value of the KLD with respect to the distribution retrieved by the { HR algorithm}.
For instance in SciND750, the flux distributions retrieved by {\em optGpSampler} have $KLD<0.05$
in $80\%$ of the cases (good agreement), $0.05 \leq KLD \leq 0.5$ in $15\%$ of them (approximate) and $KLD>0.5$ for the remaining $5\%$ (poor match). We have find that for this network {\em optSampler} gives almost systematically a better approximation than {\em gpSampler}. 
\begin{figure}[h!]
\begin{center}
\includegraphics*[width=.45\textwidth]{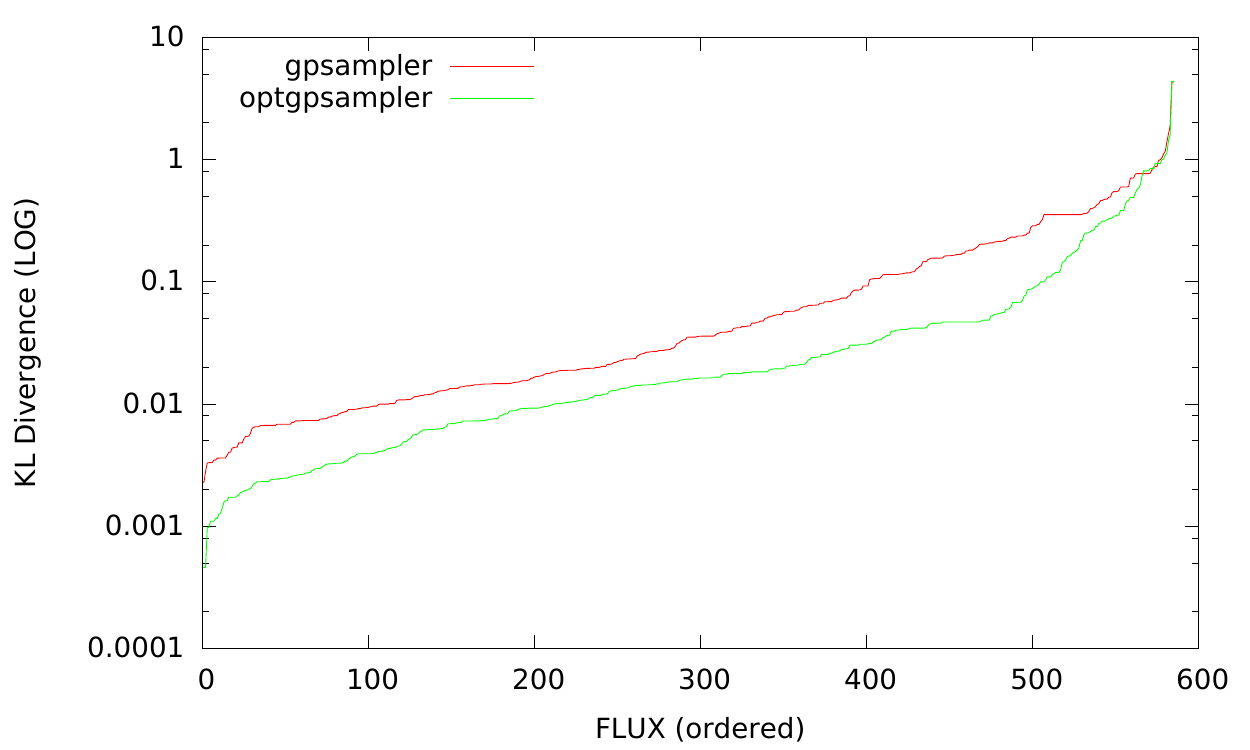}
\includegraphics*[width=.45\textwidth]{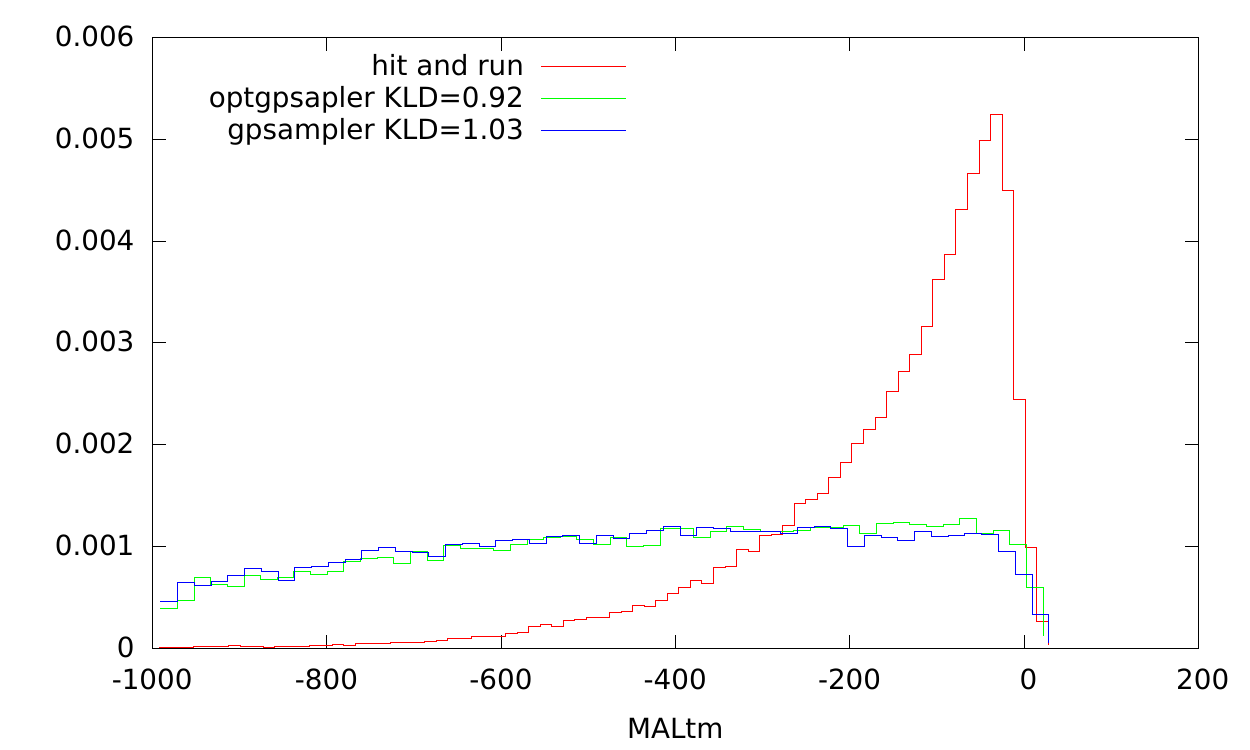}
\includegraphics*[width=.45\textwidth]{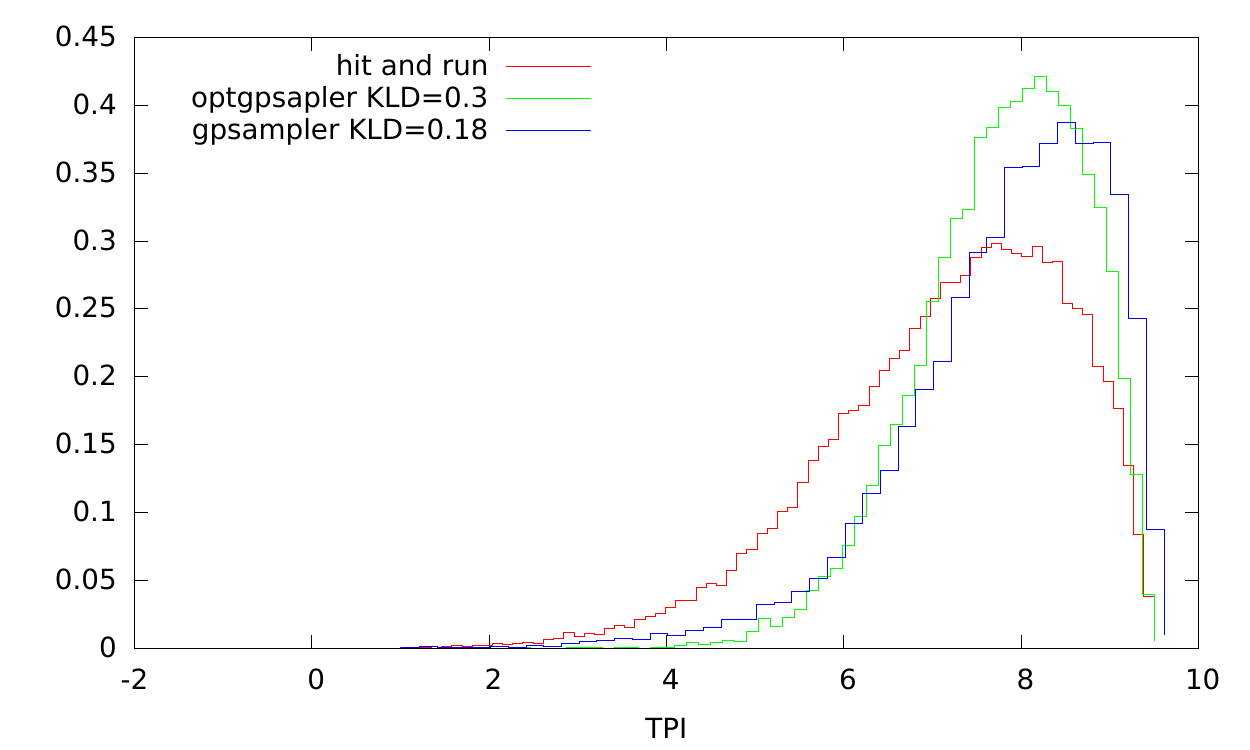}
\includegraphics*[width=.45\textwidth]{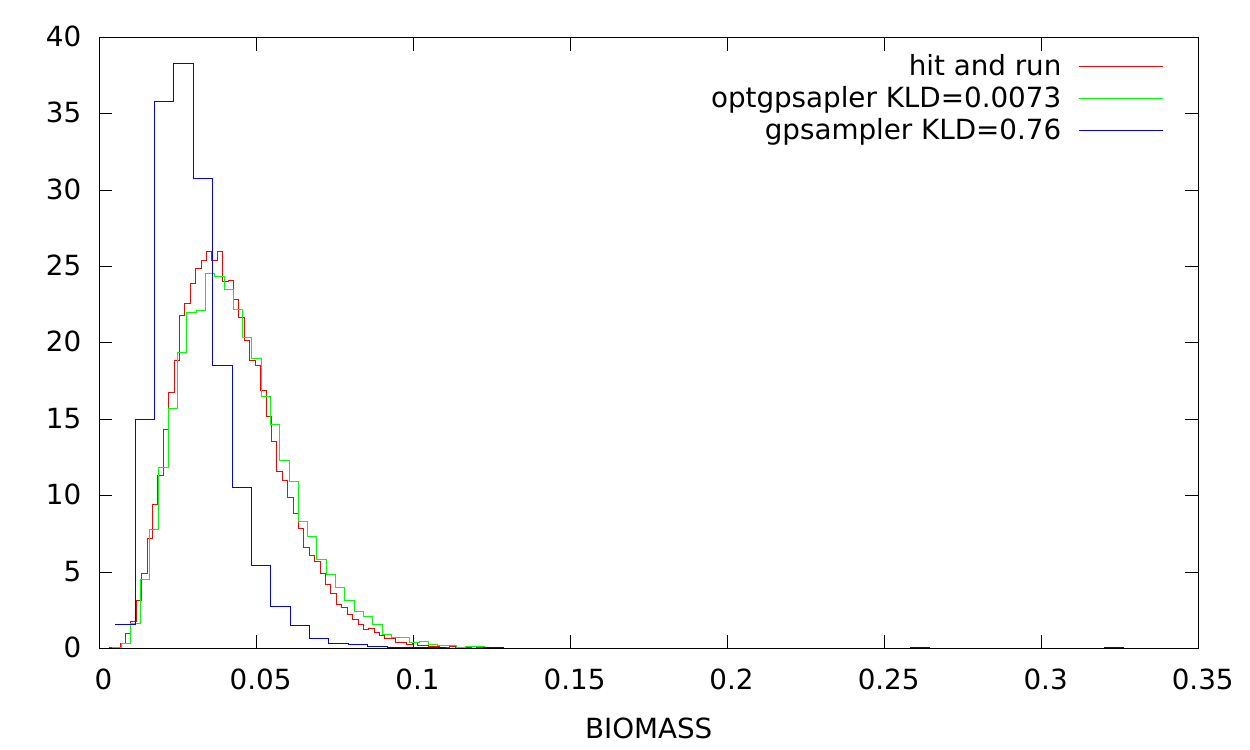}
\caption{Consistency test of {\em gpSampler} and {\em optGpSampler} with the { hit--and--run} on the model SciND750. Top left: KL divergence values of the marginal distributions of non null fluxes for {{\em gpSampler}} (red) and { \em optGpSampler} (green) compared with the { hit--and--run}, ordered for increasing values.  Top right: Marginal distributions of MALtm, this is a case in which $KLD>0.5$ ($5\%$ of the cases for { \em optGpSampler}). Bottom left: Marginal distributions of TPI, this is a case in which $0.05\leq KLD \leq 0.5$ ($15\%$ of the cases for { \em optGpSampler}). Bottom right: Marginal distributions of the growth rate, this is a case in which for { \em optGpSampler} $KLD<0.05$ ($80\%$ of the cases for { \em optGpSampler}). Histograms are obtained from $2\cdot 10^4$ points.}
\end{center}
\end{figure}
We show in {Fig.~4} the KLD values of the marginal distributions of non null fluxes for {\em gpSampler}  and {\em optGpSampler} compared with our { hit--and--run implementation}, ordered for increasing values, and some histograms representative of the aforementioned levels of approximation. For the largest network analyzed of Cervix squamous epithelial cells the level of approximation is worse, we refer to the supplementary materials. Finally, in order to test the consistency of ACHR methods on fully controlled instances we focused on sampling points from simple hypercubes of increasing dimensions. In {Fig.~5} (left) we show the marginal distribution of the first coordinate retrieved by {\em optGpSampler} on hypercubes of dimension $D=50$ and $D=500$ respectively: the distribution for the second case strongly deviates from the expected flat value. In {Fig.~5} (right) we plot the average value of the KLD over all coordinates as a function of the dimension of the hypercube: there is a clear crossover around $D\simeq 100$ from a flat behavior independent on $D$  to a monotonous increase with $D$. 
\begin{figure}[h!]
\begin{center}
\includegraphics*[width=.45\textwidth]{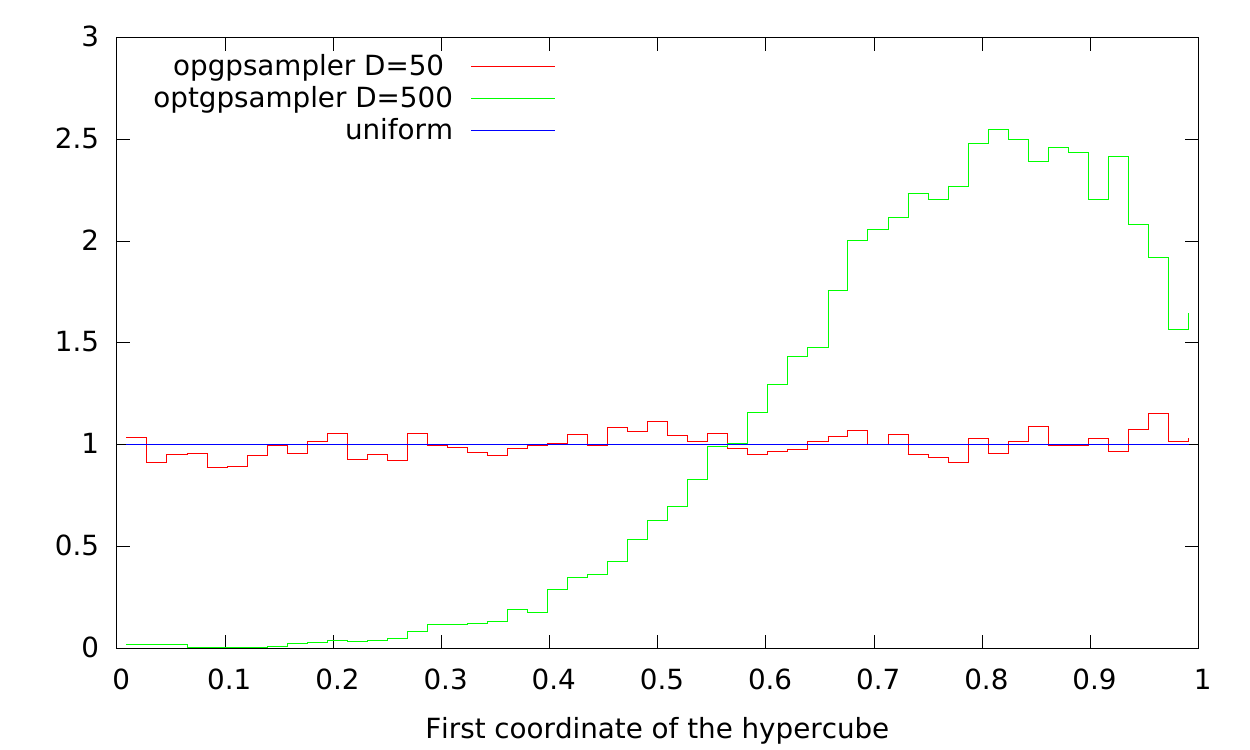}
\includegraphics*[width=.45\textwidth]{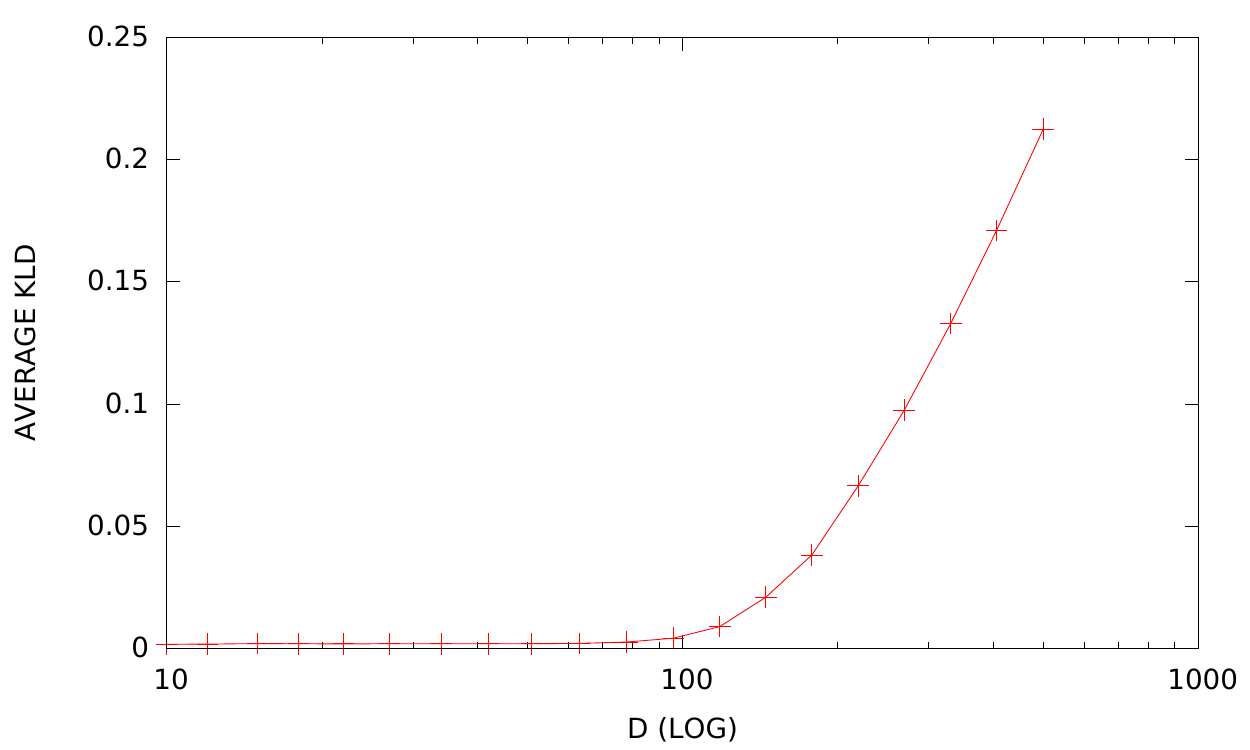}
\caption{Consistency test of {\em optGpSampler} on hypercubes, {\em gpSampler} gives similar results. Left: { Marginal} distributions of the first coordinate for { $D=50$ and 500} respectively. Right: Average value of the KLD over the coordinates with respect to the flat distribution as a function of the hypercube dimension.}
\end{center}
\end{figure}
  }
 
\section{Materials and Methods}
{
Given a $D$-dimensional convex set $P$, from which one wants to sample from, and a point inside $x_t \in P$, the standard HR algorithm is defined as follows:
\begin{enumerate}
  \item Choose a uniformly distributed direction ${\bf\theta}_t$, that is, a point extracted from the uniform distribution on the $D$-dimensional unit sphere. This can be done with the Marsaglia method, i.e. by generating $D$ independent gaussian random variables $\theta_t^i$ with zero mean and unit variance, and then normalizing the vector to unit length;
  \item Extract $\lambda^\star$ uniformly from the interval $[\lambda_{\min},\lambda_{\max}]$, where $\lambda_{\min}$ ($\lambda_{\max}$) is the minimum (maximum) value of $\lambda$ such that $\mathbf{x}_{t} + \lambda {\bf\theta}_t \in P$;
  \item Compute the point $\mathbf{x}_{t+1} = \mathbf{x}_{t} + \lambda^\star {\bf\theta}_t$, increment $t$ by one and start again.
\end{enumerate}
The starting point can be found, for instance, by interpolating between two vertices obtained by linear programming.
The second step requires to find the intersections among a line and $P$. Since $P$ is convex, the intersection points are only two, namely $\mathbf{x}_t+\lambda_{\min} {\bf\theta}_t$ and $\mathbf{x}_t+\lambda_{\max} {\bf\theta}_t$. Clearly, in order to perform the HR dynamics we should always use a {\em full-dimensional} representation of the convex set (see the supporting materials for further details); if not, $\lambda^\star=\lambda_{\min}=\lambda_{\max}$ for almost all ${\bf\theta}_t$, and dynamics is frozen.  

The decorrelation properties of the standard HR dynamics can be greately improved by a slight modification of step 1, that is, extracting ${\bf\theta}_t$ from the surface of the matching ellipsoid instead of the unit sphere. This can be easily done by multiplying a random point on the unit sphere by the symmetric matrix which defines the ellipsoid, and normalizing the resulting vector to unit length. Below we describe three different methods (illustrated in Fig. 1) in order to find or approximate the matching ellipsoid. We refer to the supposting materials for further details on the dynamics and the construction of the ellipsoid.
}
\subsection{Building the ellipsoid with { PCA}}
If we had already solved the problem, that means we have a set of uniformly distributed independent points inside the polytope, we can use them to build a matching ellipsoid by { Principal Component Analysis (PCA)}. The idea is that any sampling attempt of the polytope, even if not-equilibrating, it gathers some information on the form of the space. The connected covariance matrix from this sampling can be diagonalized and its eigenvalues and eigenvectors would give the axis of an ellipsoid that approximately matches the underlying space, the closer the nearer the sampling to equilibrium, in essence: 
\begin{itemize}
\item Perform an { HR} markov chain up to time $T$, { computing the covariance matrix of the sampled points}.
\item Diagonalize the connected covariance matrix and build an ellipsoid with axis along the principal components.
\item Use the ellipsoid for the subsequent sampling.
\end{itemize}   
The drawback of this procedure relies in the fact that ideally the sampling times $T$ should be such that the covariance matrix attains stationarity and this convergence is slow if some preprocessing is lacking. We will see that for practical purposes PCA can be used to refine the results obtained by the more direct approaches that we describe in next sections. 
 
\subsection{Building the ellipsoid with { LP}}
If we would be able to calculate the diameter of the space, then find the diameter in the orthogonal space with respect to the previous diameter and so on, we would have a matching ellipsoid whose axis coincides with such diameters.
Unfortunately, the calculation of the diameter of a convex closed space is a very hard task (think to a randomly tilted hyper-rectangle) but we can recur to an approximation by performing what is called a {Flux Variability Analysis}\cite{FVA} (FVA) in the field of metabolic network analysis. This consists in calculating the minimum and maximum values of each variable and this is a linear programming problem:
\begin{eqnarray}\label{eq4}
\textrm{Minimize/Maximize} \quad  v_i \nonumber \\
\mathbf{S \cdot v}=0, \nonumber \\
v_r \in [v_{r}^{{\rm min}},v_{r}^{{\rm max}}]
\end{eqnarray} 
that can be efficiently solved for instance with the simplex algorithm or by conjugate gradients methods. If we consider the vectors that go from the minimum to the maximum vertex for each variable, we take the vector of maximum lenght as the main axis of our ellipsoid and repeat FVA in the space orthogonal with respect to previous found axis, in synthesis:
\begin{itemize}
\item INPUT: The polytope P, a set of axis $U = \{ \mathbf{u}_1,...,\mathbf{u}_k\}$ for the ellipsoid E
\item Perform FVA within the polytope P in the space orthogonal to the subspace generated by U. 
Take the vector $\mathbf{v}$ of maximum lenght connecting min and max vertices orthogonal to U. 
\item { OUTPUT}: $\mathbf{v}=\mathbf{u}_{k+1}$ is a new axis for the ellipsoid E.
\end{itemize}  
The good point of this procedure is that it is based on the resolution of well defined set of LP problems.  
Even if this procedure is polynomial and feasible, we have to solve a large number of linear programming problems (order {$\mathcal{O}(N^2)$}). 
We have thus applied fast conjugate gradient techniques\cite{CG} as we describe in the supporting materials.

\subsection{The lovasz ellipsoid method}
We want to construct a couple of concentric ellipsoids $E,E'$ matching the polytope $P$ $E'\subseteq P \subseteq E$, 
where $E'$ is obtained from $E$ shrinking by a factor {$\mathcal{O}(1/D^{3/2})$}. This is called weak {Loewner--John} pair.
{
We define a series of enclosing ellipsoids $E_k$, 
starting with $E_0$ as the sphere with center in the origin and radius $R$ large enough in order to inscribe the body, according to the following lines:
\begin{itemize}
\item INPUT: An ellipsoid $E_k$ with its center ${\mathbf x}_k$
\item Check if ${\mathbf x}_k\in P$, if yes go to $2$, if no go to $1$
\item 1) Consider an hyperplane separating ${\mathbf x}_k$ and $P$, 
and the halfspace enclosing $P$, calculate the ellipsoid of minimal volume enclosing $H \cap E_k$ {go to} OUTPUT 1  
\item 2) Determine the endpoints of the axis of $E_k$, shrink the ellipsoid and  check if the shrinked ellipsoid $E_k'$ is inside.
if yes go to {OUTPUT} 2, if no {go to} $3$ 
\item 3) Consider an endpoint of an axis of the shrinked ellipsoid outside $P$, e.g. ${\mathbf x}_k'$, consider an separating ${\mathbf x}_k'$ and $P$,  
and the halfspace  enclosing $P$, calculate the ellipsoid of minimal volume enclosing $H \cap E_k'$ {go to} OUTPUT 1 
\item {OUTPUT} 1: A new ellipsoid $E_{k+1}$ of lower volume with center ${\mathbf x}_{k+1}$, update $k$, repeat from INPUT.
\item {OUTPUT} 2: A weak Loewner-John {ellipsoid.}
\end{itemize}
This algorithm is substantially an expanded version 
of the famous ellipsoid method used to demonstrate the feasibility of linear programming problem.
Upon calculating the reduction in volume of the enclosing ellipsoid after one step, it can be demonstrated that this series converges in polynomial time to a weak {Loewner--John} pair.
We refer to \cite{lovaszbook,survey} for further details.

\section{Conclusions}
In this article we have proposed rounding methods in order to reduce the condition number for the application of the { hit--and--run (HR) Markov Chain Monte Carlo} to the problem of the uniform sampling of steady states in metabolic network models. 
They are based on matching the polytope under exam with an ellipsoid that can be used to { bias the HR random walker, still sampling the flux space in a uniform way}.
Such ellipsoids were built by applying principle component analysis to previous sampling, by solving a set of linear programming problems {--} similarly to a technique called Flux Variability Analysis in the field of metabolic network analysis, and by the lovasz ellispoid method. In particular the last two can be calculated in polynomial times.
We have applied them in order to sample the feasible steady state of three  metabolic network reconstruction of growing size where we successfully removed the ill{--}conditioning and reduced the sampling times of a  factor $O(10^{7-12})$ with respect to the normal { HR} dynamics. 
The lovasz method or the LP method alone were sufficient to remove the ill{--}conditioning. With our implementation the first gives better results on the small network, the second on the genome scale networks, whereas the PCA can be used to refine the results of the other two, since in this case it is possible to obtain the ellipsoid from a diagonalization of a good estimator of the stationary connected covariance matrix. The overall procedure, preprocessing and subsequent sampling is feasible in genome scale networks, in agreement with theoretical results on the computational complexity regarding these tasks. The rounding preprocessing is still quite intensive on large genome scale models with our implementation and this leaves space for optimizing time performances that we leave for further investigations.
Even if there could be faster methods in order to sample points inside convex polytopes, the { HR Monte Carlo} is guaranteed to converge to an uniform distribution. It could be used thus in order to test the correctness of fast message-passing \cite{MPperez} or ACHR-based algorithms in their convergence to an uniform distribution.
We have thus compared the samples retrieved by the { HR method} with two ACHR based method provided with the COBRA toolbox {\em gpSampler} and {\em optGpSampler}. 
We checked that they generate similar distributions with rather different speed, {\em optGpSampler} being much faster.
We have found that in the small network these ACHR-based methods are consistent with the uniform sampling  provided by the hit an run according to the Kolmogorov-Smirnov test. 
On genome scale networks the flux distributions retrieved by these methods do not pass the KS test, but give an approximation that we quantified by calculating their Kullback Leibler divergence with respect to the distribution { obtained with HR dynamics}. 
In some cases we detected inconsistencies that get worse on higher dimensions. This behavior has been highlighted upon sampling high dimensional hypercubes. We want to mention finally that in regard to the problem of sampling the steady states of a metabolic networks a rigorous implementation of thermodynamic constraints possibly renders the space non-convex\cite{deMartino12}. The development of Montecarlo methods for the sampling of non convex spaces is a difficult open issue. The { HR} algorithm  applied to the sampling of non--convex bodies is not guaranteed to converge in polynomial times, an aspect that needs further investigations.
}
\section*{Acknowledgments}
DDM thanks F.Capuani and A. De Martino for interesting discussions.  
This work is supported by the DREAM Seed Project of the Italian Institute of Technology (IIT). The IIT Platform Computation is gratefully acknowledged.

\section*{Appendix}
\subsection*{A The dimension of the polytope and the search for a starting point inside}
In order to run any hit and run dynamics, we need a full--dimensional representation of the flux space.
It could be that the constraints defining the polytope are infeasible: i.e. the inequalities cannot be satisfied.
Furthermore it could be that some variables attain their bounds: 
this would reduce the dimension of the polytope whose upper bound is the dimension of the kernel of the stoichiometric matrix.
We can check the feasibility of the space, calculate the dimension of the polytope, find the right variables and a point inside with the use of relaxation algorithms.
Upon finding a kernel basis (the columns of the orthogonal matrix $B_{ik}$) of the Stoichiometric matrix we can write
\begin{equation}
v_i = \sum_{k=1}^K B_{ik} u_k
\end{equation}
then we have the system of inequalities 
\begin{equation}\label{primal}
v_{r}^{{\rm min}} \leq  \sum_{k=1}^K B_{rk} u_k \leq v_{r}^{{\rm max}}.
\end{equation}
This system has the form $\mathbf{A \cdot u} \leq \mathbf{b}$ , we will refer to this notation for simplicity in the following.
A relaxation algorithm(\cite{shrij}) works in the following way: starting from any point
\begin{itemize}
\item Calculate the less satisfied constraint $ i_0 = \textrm{min}_i ( \mathbf{b}- \mathbf{A \cdot u} )_i$
\item If this is positive the point is inside (EXIT) else update in the orthogonal direction $u_k = u_k - \alpha A_{i_0 k}$ and start again 
\end{itemize}  
If a solution exists with {\em strict inequalities} the polytope is full-dimensional, 
the relaxation algorithm converges in polynomial time and from the resulting point we can start the hit-and-run dynamics.
Otherwise the dual theorems of the alternative\cite{shrij} tells us that there is a non-trivial solution $\mathbf{w}$ to the system
\begin{eqnarray}\label{dual}
\mathbf{w \cdot A }=0 \nonumber \\
\mathbf{w \cdot b}<0 \nonumber \\
\mathbf{w}\geq 0
\end{eqnarray}
In this case the constraints of the primal system (\ref{primal}) that corresponds to the non-zero variables of solutions of the dual system (\ref{dual}) attain their bounds, 
i.e. they count as equations and the dimension of the polytope is reduced. 
For practical purposes, the polytopes defined by the set of feasible steady states of metabolic networks are almost full-dimensional: few variables attain their bounds
and they can be found by coupling relaxation algorithms to an exhaustive search within less unsatisfied constraints. 
Alternatively, a point inside can be found by the ellipsoid method that we describe within the implementation of the lovasz method. 

\subsection*{B Using the ellipsoid in order to eliminate the ill-conditioning}
  Once we have a matching ellipsoid this can be used to define an affine transformation 
  that reduces considerably the condition number of the resulting sampling space.
  However,  in practical cases, the matrix of constraints can be sparse, a property that could be lost during the transformation.
  This is the case of metabolic networks, in which a reaction usually involves $2-6$ compounds. 
  We have found by parsing that it is faster in practice to select a direction by taking a point 
  at random from the surface of the ellipsoid rather than transforming the space. If we tranform the space
  the problem of finding the extrema of the segment (step 2 of the hit-and-run algorithm) is rather time-consuming.  
  The results of the sampling are the same but the machine time per 
  Monte Carlo step is approximatively $30\%$ lower for the polytope under exam. 
\subsection*{C A conjugate gradients method for linear programming}
Conjugate gradient methods are widely used algorithms in order to minimize non-linear functions,
in particular for quadratic functions they show linear convergence. In this last case they basically construct
at each step vectors of a conjugate basis, that are used thereafter as search directions. In this way it is possible to avoid the repetitions that would be present applying a steepest descent.
We use the method developed in \cite{CG}  to solve linear programming, of wich we recall the main steps.  
We want to solve
\begin{eqnarray}\label{CGLP}
\textrm{Minimize} \quad  \mathbf{c \cdot u} \nonumber \\
v_{r}^{{\rm m}} \leq  \sum_{k=1}^K B_{rk} u_k \leq v_{r}^{{\rm M}}.
\end{eqnarray}
Let's consider the minimum of the piece-wise quadratic function
\begin{eqnarray}
F_{\sigma}(\mathbf{u}) = \mathbf{c \cdot u} +\nonumber \\
+\sigma \sum_r \theta( (\mathbf{B u})_r -  v_{r}^{{\rm M}}) (  (\mathbf{B u})_r -  v_{r}^{{\rm M}}  )^2   + \theta( v_{r}^{{\rm m}}  -(\mathbf{B u})_r) (  (\mathbf{B u})_r -  v_{r}^{{\rm m}}  )^2 
\end{eqnarray}
where $\theta$ is the Heavyside step function that enforces the constraints whose relative strenght is parametrized by $\sigma$.
The minimum  ${\bf u}_{min}(\sigma)$ is an approximate solution of the linear programming problem, whose approximation is controlled by $\sigma$.
For practical purposes it is convenient to regard ${\bf u}_{min}(\sigma)$ as linear function of $1/\sigma$, such that the limit value $\sigma \to \infty$ can be extrapolated from two points. 
In each analytic sector this function is quadratic such that the method has good and controlled properties of convergence. 
From its definition $F$ is continuous with its first-order derivatives (that are piece-wise linear).
The conjugate gradients are then calculated along the following lines: 
\begin{itemize}
\item Input: a point $\mathbf{u}_t$, a direction $\mathbf{d}_t$.
\item 1) Calculate the minimum over $\alpha \in\mathbb{R}$ of $F(\mathbf{u}_t+\alpha \mathbf{d}_t)$.
\item 2) Update the point $\mathbf{u}_{t+1} =  \mathbf{u}_t+\alpha \mathbf{d}_t$, calculate the gradient of F here $\mathbf{g}_{t+1} = \nabla F(\mathbf{u}_{t+1})$.
\item 3) Update direction, $\beta = \frac{\| \mathbf{g}_{t+1} \|^2}{ \| \mathbf{g}_{t} \|^2  }$, $\mathbf{d}_{t+1} = -\mathbf{g}_{t+1}+\beta \mathbf{d}_t $.
\item Output: a point $\mathbf{u}_{t+1}$, a direction $\mathbf{d}_{t+1}$.
\end{itemize} 
We start from the the origin and the gradient of $F$ there, and we end when the norm of the gradient is small enough (less than $\epsilon=10^{-3}$).
The line search (step 1) can be performed by keeping attention to the piece-wise continuos and monotonous form of $\frac{dF}{d\alpha}$. 
We refer to \cite{CG} for further details.

\subsection*{D Example: strongly heterogeneous hyper-rectangles}
As a matter of illustration we shall show the application of the methods to a control set of polytopes: the hyper-rectangles of dimension $D$ with edges parallel to the axis coordinates and of exponentially increasing size:
\begin{equation}
0 \leq x_i \leq 2^{i-1}  \qquad \forall i\in 1 \dots D 
\end{equation}
The mixing time scales as $\tau \propto 2^{D-1}$ for the simple hit-and-run dynamics, i.e. in this extreme case the ill-conditioning is so strong that the convergence time of the algorithm looses its polynomial scaling  with the dimensions if preprocessing is lacking.
If we build the ellipsoid with the first method the situation improves but still we would have to wait a long time in order to fully perform the preprocessing.
On the other hand the LP and lovasz method remove completely the ill-conditioning in one shot.
In particular the LP method converges immediatly given the simple nature of the linear programming problems under exam while the lovasz method it takes around $5$s with our implementation.
As we can see from fig.1 the integrated autocorrelation times of the coordinate axis in $D=20$ are all equal for the dynamics with the ellipsoid based on LP and lovasz method,
on the other hand the ellipsoids built with PCA can remove the ill-conditioning up to a certain degree that depends on the preprocessing time.
\begin{figure}[h!]\label{fig2}
\begin{center}
\includegraphics*[width=.6\textwidth,angle=0]{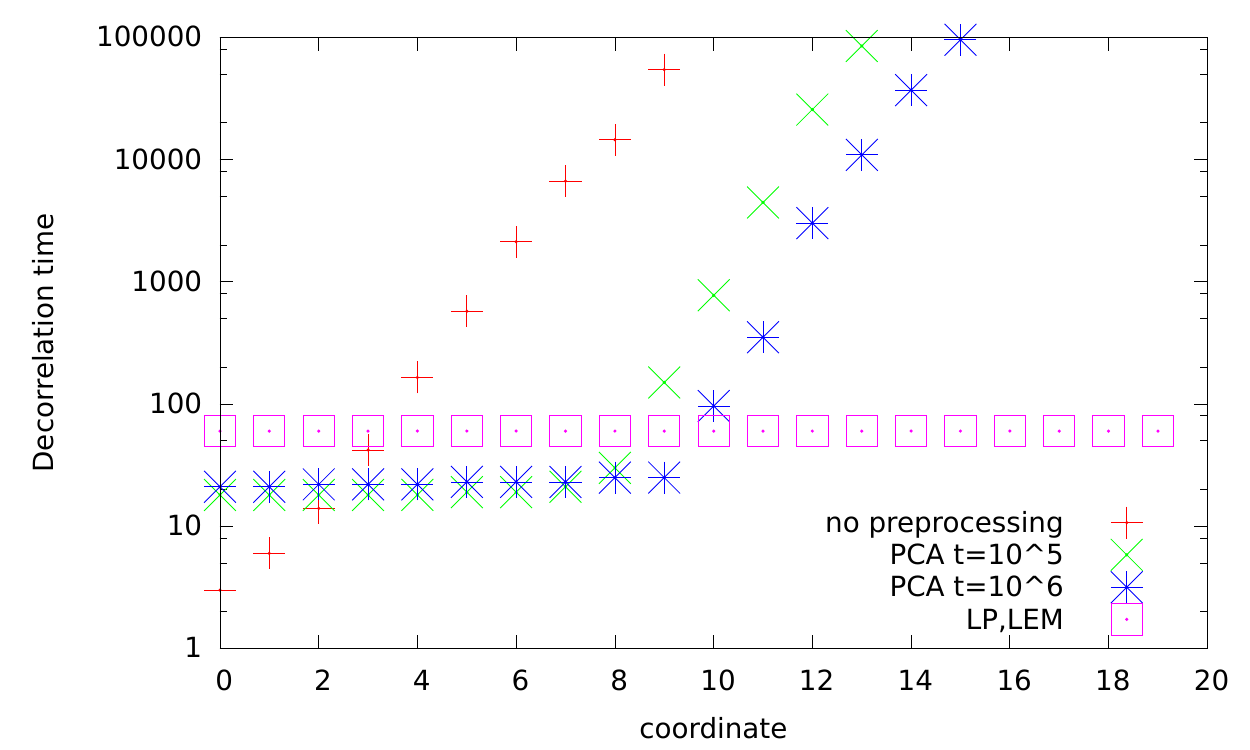}
\caption{integrated autocorrelation times of the coordinate axis of a highly heterogeneous hyperrectangle in $D=20$ sampled with hit-and-run dynamics with and without preprocessing.}
\end{center}
\end{figure}

\subsection*{E Integrated autocorrelation times}
Suppose we have a stationary, discrete, time series $x_t$, upto time $T$.
We want to estimate the time  after which we can consider the points uncorrelated.
 If we define the correlation function $C(t) = \langle x_t x_0 \rangle - \langle x \rangle^2$, upon calculating the variance of the mean upto time $T$ we have
 \begin{equation}
 \sigma^2(\bar{x}) = \frac{\sigma^2(x)}{T} \left[ 1+2 \sum_{t=1}^{t=T} \left( 1-\frac{t}{T} \frac{C(t)}{\sigma^2(x)} \right)  \right]
 \end{equation}
 Upon comparing with the usual expression for uncorrelated points we can see that the factor
 $\langle \tau_{int}^T \rangle= 1+2 \sum_{t=1}^{t=T} \bigl(1-\frac{t}{T} \frac{C(t)}{\sigma^2(x)} \bigr)$ is the quantity of interest, in particular its limit $T\to \infty$, 
 that is the integrated autocorrelation time. 
  Unfortunatively the variance of the estimator of this quantity usually diverges for correlated variables. 
This is our case as well, as it can be seen from fig. below
\begin{figure}[h!]\label{figint}
\begin{center}
\includegraphics*[width=.6\textwidth,angle=0]{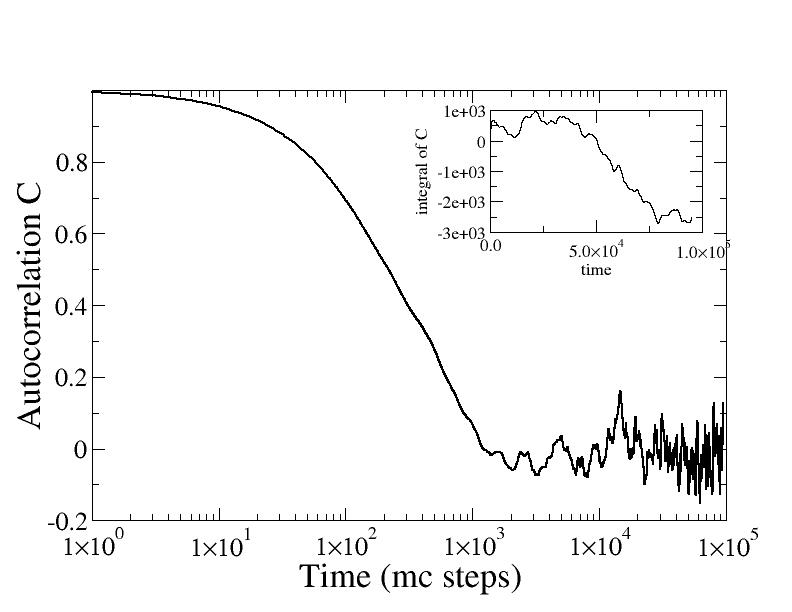}
\caption{Autocorrelation function during ellipsoid-based hit-and-run markov chain Montecarlo of one reaction flux in E.Coli Core 
calculated averaging over $2 \cdot 10^5$ points. 
The signal-to-noise ratio decreases  strongly when the function approaches zero, leading to a difficult numerical estimate of its integral (integrated autocorrelation time, inset).}
\end{center}
\end{figure}
\begin{figure}[h!!!!!]\label{figint2}
\begin{center}
\includegraphics*[width=.6\textwidth,angle=0]{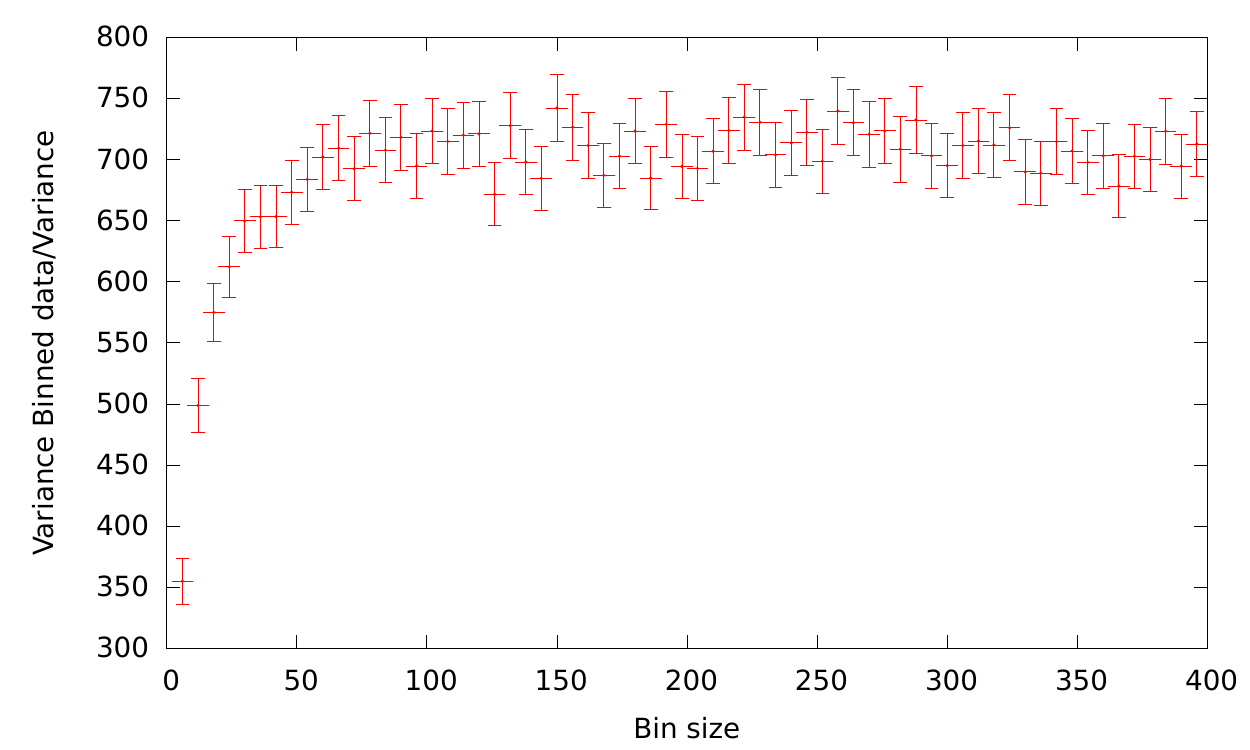}
\caption{Estimate of the integrated autocorrelation time of the function depicted in fig. 2 by binning the data. 
X axis: Bin lenght; Y axis: ratio of the variance of the binned data over the variance of unbinned data; error bars calculated from a gaussian approximation.}
\end{center}
\end{figure}
  A common procedure in order to have an estimate of the integrated autocorrelation time is to  bin the data, i.e. we divide $T$ into $N_b$ bins each of size $K$ and we ``renormalize'' the data, i.e. we define
  \begin{equation}
  x_{i}^K = \frac{1}{K} \sum_{t=iK}^{(i+1)K-1} x_t
  \end{equation}
  We calculate the variance of this renormalized data and consider the ratio with respect to the unbinned variance. This is for large enough $K$ a good estimate of the integrated autocorrelation time. 
In fig.8 we report the ratio of the variances as a function of the bin size for the autocorrelation function depicted in fig. 7 

This tends to a plateau that is the integrated autocorrelation time. 
Regarding the problem of transient phenomena related to the choice of the initial point we finally  point out that 
for the case  in which we implement a successful preprocessing, 
the size of the sample over which we perform our estimate is so large that any bias due to transients
is completely leveraged out.
\subsection*{F Waiting for the effective convergence of em gpSampler}
\begin{figure}[h!]
\begin{center}
\includegraphics*[width=.6\textwidth,angle=0]{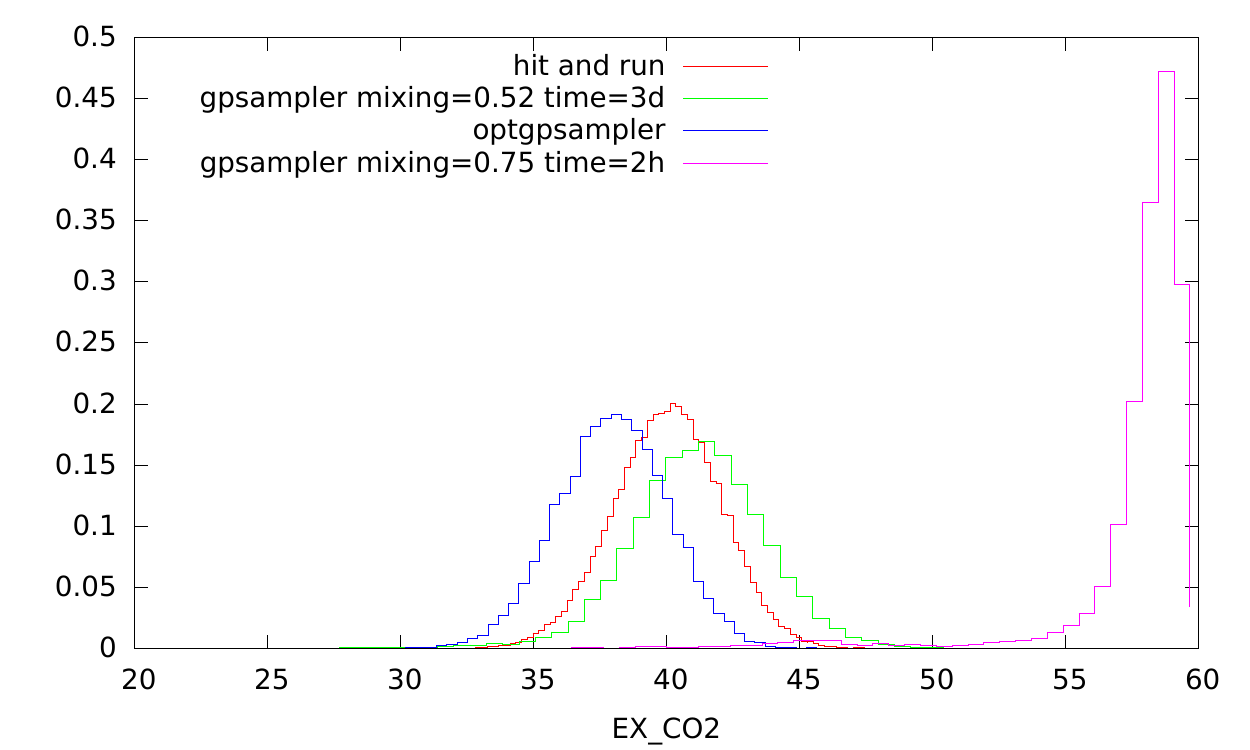}
\caption{Histograms of the carbon dioxide excretion in SciND750 retrieved by the hit and run, {\em optGpSampler} and {\em gpSampler}, the latter upon waiting different times for convergence.}
\end{center}
\end{figure}
It has been reported in \cite{2014optGpSampler} that the artificially centered based methods {\em gpSampler } and {\em optGpSampler} could give different results for the same network model. We were able to reproduce these kind of results but we have found that eventually {\em gpSampler} gives different and better results upon waiting until the mixing reaches at least the value $0.55$. This is reported in fig 4 where we show the histograms of the carbon dioxide excretion in SciND750 retrieved by the hit and run, {\em optGpSampler} and {\em gpSampler}, the latter upon waiting different times for convergence.
Upon waiting some hours the result is similar to the one reported
in  \cite{2014optGpSampler}, where it seems that it misses completely to match the value of the hit and run that is instead in good agreement with {\em optGpSampler}.
Upon waiting until the mixing fraction reaches a value lower than $0.55$ we have found a much better result that we report in fig. 9.   Thus we have checked that the two ACHR methods give similar distributions upon waiting an effective convergence of {\em gpSampler}, that for the largest network analyzed requires around one week of machine time.
\subsection*{G Worsening of the consistency of the ACHR methods with the hit and run for increasing dimensions}
We observe that the consistency of ACHR methods with the results of the hit and run gets worse for higher dimensions.
In addition to the hypercube case shown in the main text we report here  in fig 5 the Kullback-Leibler divergences of the marginal flux distributions obtained with {\em optGpSampler} with respect to the hit and run. The values of the KLD are sorted in increasing order for the three metabolic networks examined. {\em gpSampler} gives slightly worse but similar results.    
\begin{figure}[h!]
\begin{center}
\includegraphics*[width=.6\textwidth,angle=0]{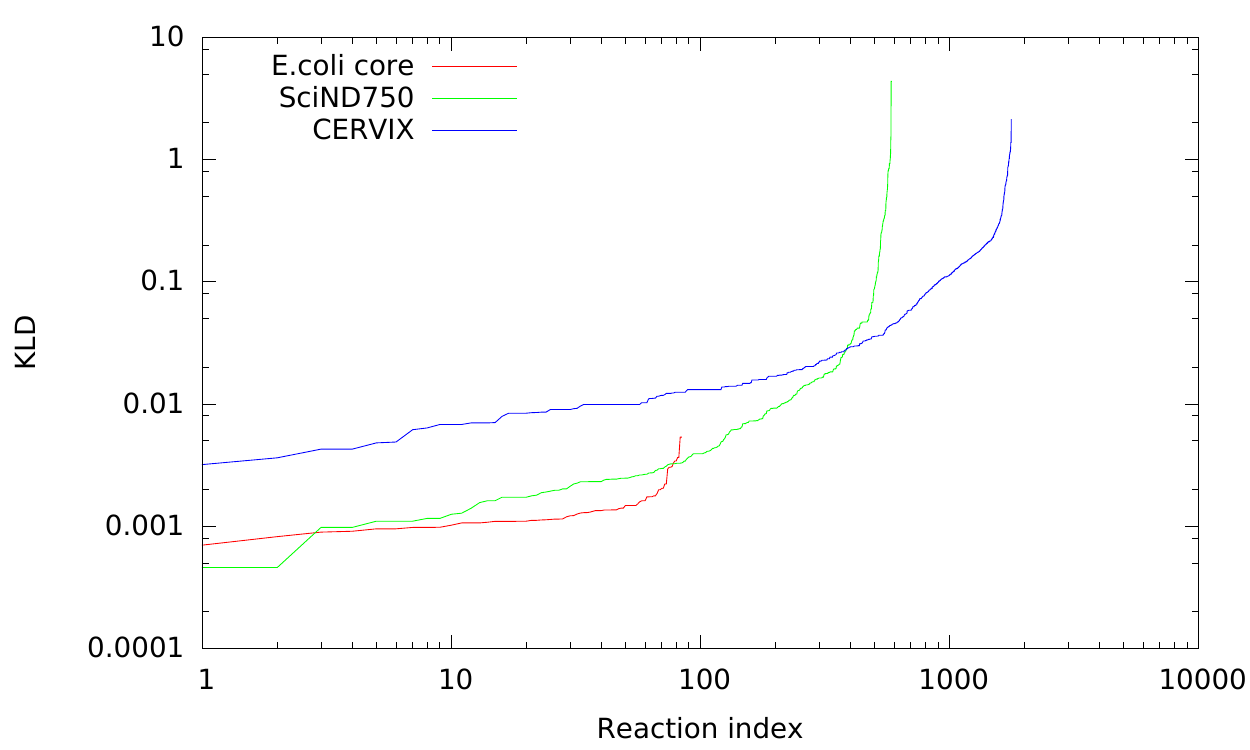}
\caption{Kullback-Leibler divergences of the marginal flux distributions obtained with {\em optGpSampler} with respect to the hit and run ordered for increasing values for the three metabolic networks examined.}
\end{center}
\end{figure}

\section*{References}
\bibliographystyle{unsrt}
\bibliography{refplos}

\end{document}